\documentclass[8.5pt,twoside,twocolumn]{article}
\usepackage[super,sort&compress,comma]{natbib} 
\usepackage[version=3]{mhchem} 
\usepackage{times,mathptmx}
\usepackage{sectsty}
\usepackage{balance} 
\usepackage{graphicx} 
\usepackage{lastpage}
\usepackage[format=plain,justification=raggedright,singlelinecheck=false,font=small,labelfont=bf,labelsep=space]{caption} 
\usepackage{fancyhdr}
\usepackage{bm}
\usepackage{subfigure}
\newcommand{\rcut}{r_{\mathrm{cut}}}

\newcommand{\es}{\textsf{ESPResSo}}
\newcommand{\mmu}{\mu^{2}}
\newcommand{\myfigurewidth}{1.0\columnwidth}
\newcommand{\mypagewidth}{1.6\columnwidth}
\bmdefine\bmu{\mu}
\bmdefine\bet{u}
\bmdefine\btau{\tau}
\bmdefine\bomega{\omega}
\bmdefine\bmeta{k}
\bmdefine\brho{\rho}
\bmdefine\bn{n}
\bmdefine\bmm{m}
\bmdefine\bk{k}
\bmdefine\br{r}
\bmdefine\bx{x}
\bmdefine\bxi{\xi}
\bmdefine\bT{T}
\bmdefine\bF{F}
\bmdefine\bv{v}
\bmdefine\bW{W}
\bmdefine\bI{I}
\bmdefine\bl{l}

\newcommand{\ie}{\emph{i.e.}~~} 
\newcommand{\eg}{\emph{e.g.}~~}

\begin{document}

\thispagestyle{plain}
\fancypagestyle{plain}{
\renewcommand{\headrulewidth}{1pt}}
\renewcommand{\thefootnote}{\fnsymbol{footnote}}
\renewcommand\footnoterule{\vspace*{1pt}%
\hrule width 3.4in height 0.4pt \vspace*{5pt}} 
\setcounter{secnumdepth}{5}

\makeatletter 
\def\subsubsection{\@startsection{subsubsection}{3}{10pt}{-1.25ex plus -1ex minus -.1ex}{0ex plus 0ex}{\normalsize\bf}} 
\def\paragraph{\@startsection{paragraph}{4}{10pt}{-1.25ex plus -1ex minus -.1ex}{0ex plus 0ex}{\normalsize\textit}} 
\renewcommand\@biblabel[1]{#1}            
\renewcommand\@makefntext[1]%
{\noindent\makebox[0pt][r]{\@thefnmark\,}#1}
\makeatother 
\renewcommand{\figurename}{\small{Fig.}~}
\sectionfont{\large}
\subsectionfont{\normalsize} 

\fancyfoot{}
\fancyfoot[RO]{\footnotesize{\sffamily{~\textbar  \hspace{2pt}\thepage}}}
\fancyfoot[LE]{\footnotesize{\sffamily{\thepage~\textbar}}}
\fancyhead{}
\renewcommand{\headrulewidth}{1pt} 
\renewcommand{\footrulewidth}{1pt}
\setlength{\arrayrulewidth}{1pt}
\setlength{\columnsep}{6.5mm}
\setlength\bibsep{1pt}

\twocolumn[
  \begin{@twocolumnfalse}
\noindent\LARGE{\textbf{Phase diagram for a single flexible Stockmayer polymer at zero field}}
\vspace{0.6cm}

\noindent\large{\textbf{ Joan J. Cerd\`{a},$^{\ast}$\textit{$^{a\ddag}$} Pedro A. S\'anchez,\textit{$^{b}$} Christian Holm,\textit{$^{b}$}  and Tom\`{a}s Sintes \textit{$^{a}$}}}\vspace{0.5cm}

\vspace{0.6cm}

\noindent \normalsize{The equilibrium conformations of a flexible permanent magnetic chain that consists of a sequence of linked magnetic colloidal nanoparticles with short-ranged Lennard-Jones attractive interactions (Stockmayer polymer) are thoroughly analysed via Langevin dynamics simulations. A tentative phase diagram is presented for a chain of length $N=100$. The phase diagram exhibits several unusual conformational phases when compared with the non-magnetic chains. These phases are characterised by a large degree of conformational anisotropy, and consist of closed chains, helicoidal-like states, partially collapsed states, and very compact disordered states. The phase diagram contains several interesting features like the existence of at least two 'triple points'.
}
\vspace{0.5cm}
 \end{@twocolumnfalse}
  ]

\section{Introduction}



\footnotetext{\textit{$^{a}$~Instituto de F\'{\i}sica Interdisciplinar y Sistemas Complejos, IFISC (CSIC-UIB). Universitat de les Illes Balears. E-07122 Palma de Mallorca, Spain.}}
\footnotetext{\textit{$^{b}$~Institut f\"{u}r Computerphysik. Universit\"at Stuttgart. 70569 Stuttgart, Germany.}}


\footnotetext{\ddag ~E-mail: joan@ifisc.uib-csic.es}


Artificial magnetic filaments can be obtained by permanently linking magnetic colloids to form a chain. These magnetic chains represent the equivalent to magnetic polymers but at a supra-molecular scale. Whereas magnetic polymers keep their magnetic properties only at T$<$100K, \cite{2002-kamachi, 2004-blundell} magnetic filaments can retain their magnetism at zero field and at room temperature if the size of the nanocolloids is chosen adequately.

The path towards the synthesis of such permanent magnetic filaments has been possible thanks to a progressive increase in the abilities to control the size of magnetic colloids and the nature of bonds  between the colloidal particles \cite{2011-sanchez,2011-wang,2007-martinez-pedrero, 2008-martinez-pedrero,2008-liu,2007-evans,2011-benkoski,2012-breidenich}. Among all studies we note the very recent work of Sarkar and Mandal \cite{2012-sarkar} who have performed the synthesis of magnetic chains  using DNA as a template on which they have directly grown the magnetic nanoparticles with sizes ranging between $7$ and $17$ nanometers. It is also worthwhile to point out that there have been successful attempts by Zhou et al. to lock and preserve the structural conformations of filaments made of magnetic cobalt nanocolloids of $20nm$ in size\cite{2009-zhou}. Goubault\cite{2005-goubault} et al. have achieved the synthesis of flexible magnetic filaments by the simple procedure of bridging the surfactant layers carried by ferrofluid particles adsorbed on top of their surface in the presence of an external magnetic field. Once the bridging has occurred, the particles are irreversibly linked and the external magnetic field can be removed.

The growing interest in the relatively new field of magnetic filaments is driven by the promising novel technological applications. They can be thought as improved substitutes for current ferrofluids, as the new elements for magnetic memories, as chemical and pressure nanosensors, or have useful applications for medical purposes, to mention just a few.\cite{2011-wang} The use of non magnetic polymer colloids in ferrofluids that behave as 'inverse magnetic filaments' may also have potential applications as already shown in the assembly of non-permanent photonic crystals.\cite{2012-liu} In general, in most of the applications, the knowledge of the different types of equilibrium structures that filaments may adopt is of extreme importance. Nonetheless, to date, very little is known about the structures that permanent magnetic chains may adopt as a function of temperature, length, and other related parameters like the magnetic moment of the particles or the strength of the attractive interactions among colloidal particles.

For the case of non-magnetic attractive chains the study of their phase transitions has been exhaustive, see refs. \cite{1997-zhou,2007-vogel,2010-seaton,2009-polson,2009-schnabel-cpl,2006-opps,2005-polson}, and refs.\cite{2008-binder,2003-baysal} for a review. In the case of semiflexible attractive polymers it is known that there exist several conformational phases different from the typical swollen coils, collapsed globules, crystal and glassy states. Several studies have shown semiflexible chains to possess toroidal or disk-like phases.\cite{1998-ivanov,2008-binder,2000-ivanov,2003-stukan,2008-higuchi,2012-seaton} Helix structures have also been found for some very specific square-well potentials.29 Related to this, helical long-lived transient states have also been identified for chains with truncated Lennard-Jones potentials.\cite{2008-sabeur}

For magnetic chains the number of studies is much lower. In addition to the previously mentioned studies devoted to the experimental synthesis of magnetic filaments, there have been several attempts to obtain phase diagrams for magnetic chains using Ising or Heisenberg-like monomers in a good solvent\cite{1999-garel,2004-huang,2006-luo-jcp,2006-luo-pol,2006-huang} that correspond to the case of non-attractive colloids. Henceforth, we will refer to such kind of chains as non-sticky filaments. In addition, some studies have also dealt theoretically with the study of the magnetostatics of chains made of magnetic nanoparticles of different shapes.\cite{2011-phatak} The derivation of the partition function, the intra-chain correlations, and the coil–globule transition for flexible non-sticky magnetic chains in the limits of zero and infinitely strong external magnetic fields has also been pursued.\cite{2002-morozov} The phase diagram of homopolymers with their magnetic dipoles constrained to be locally perpendicular to the chain in order to mimic a protein has been also studied with a virial and Landau approach.\cite{1997-pitard}

A large fraction of the existing studies has been devoted to the properties of magnetic filaments made of paramagnetic or super-paramagnetic non-sticky chains. Most of these studies treat the filaments as elastic rods to use them as micro-propellers (microswimmers)\cite{2004-shcherbakov,2008-erglis-jpcm, 2008-erglis-mh, 2003-biswal, 2005-cebers, 2004-cebers, 2009-belovs-pre,2009-belovs-jpa,2011-benkoski,2009-snezhko} under the action of an external field, or as actuators\cite{2007-evans,2010-benkoski} with the purpose of performing tasks similar to those of micrometric magnetic cilia,\cite{2009-fahrni,2011-babataheri} but at the nanoscale.

For lower dimensionalities and non-sticky magnetic filaments, Sánchez et al.\cite{2011-sanchez} have recently addressed the adsorption properties and the equilibrium conformation properties on an adsorbed attractive surface via numerical simulations. In a subsequent study the same authors\cite{2013-sanchez} report the different structural regimes displayed by non-attractive flexible magnetic filaments immersed in a good solvent as a function of the relative strength of the magnetic versus thermal forces. The presence of a short-ranged Lennard-Jones attractive interaction, in addition to the point dipole, characterises the so-called Stockmayer chains, and is expected to strongly modify the behaviour of these filaments. The study of the effects of the short-ranged attraction on the equilibrium conformations of a Stockmayer polymer is the main purpose of the present work.

Relevant to our study of Stockmayer chains are studies of clusters of free particles that interact via Stockmayer potentials (Lennard-Jones plus point dipole potentials) in the limit of very low temperatures. Miller and Wales\cite{2005-miller} have found that clusters of particles interacting via Stockmayer potentials exhibit a rich variety of ground states that includes rings and different types of coils with several topological knots. The type of ground state exhibited by a cluster of Stockmayer particles was found to depend strongly on the dipole moment and the number of particles. Nonetheless, it is not clear if the conformations of a single magnetic sticky filament at low temperatures will resemble the ground states found for Stockmayer clusters that lack permanent links between particles.

In this paper we study the influence of the magnetic interactions on the phase diagram of sticky and non-sticky magnetic chains in three-dimensions via the use of extensive Langevin dynamics simulations. In Section \ref{sec2} we describe the numerical model and the details of the simulations. In Section \ref{sec3} we present and discuss the results with emphasis in the structures and phases found in the weak and strong attractive interaction regime, and a tentative phase diagram is also presented. Finally, a summary and a discussion of the conclusions are presented in Section \ref{sec4}.


\section{Numerical Model}
\label{sec2}
The magnetic filament is modelled as a bead-spring chain made of a sequence of $N$ magnetic beads (colloidal particles) of diameter $\sigma_e$, carrying a point dipole $\bmu_e$ at their centre. Henceforth, the subindex $e$ denotes the experimental values of the physical quantities we use, whereas the absence of such a subindex means the quantity is expressed in reduced units. Throughout this work, we will express the results in reduced units, thus, for a length le its corresponding reduced value is $l=l_e/\sigma_e$. In all the simulations carried out the diameter of the colloidal particles is set to $\sigma=1$, so all length scales are measured in units of this diameter. Therefore, our results are expected to be valid to any particle size as far as the magnetism of the colloidal particles can be approached by point dipoles fixed in the lattice structure of particles, and sedimentation forces are negligible.

Two different types of particles will be considered: non sticky particles, where the predominant interaction between particles is the steric repulsion due to their cores and sticky particles, in which in addition to the repulsion there exists an attractive pair-wise interaction among the particles. We will refer to this interaction as {\em 'the $LJ$ interaction'}. The attractive interaction between two particles $i$ and $j$ will be modelled via the following potential that combines the core repulsive part (cutoff $r_{cut}=2^{1/6}\sigma$), and the attractive part ($r_{cut}=2.5\sigma$),

\begin{equation}
      \label{Ushort}
      U_{att}(r) = V_{tsLJ}(r,\sigma,1,r_{cut}=2^{1/6}\sigma) + V_{tsLJ}(r,\sigma,\epsilon,r_{cut}=2.5\sigma)
   \end{equation}
where $r$ is the distance between the centres of the particles $i$ and $j$, \ie $r=| \br_i-\br_j |$, and $V_{tsLJ}$ is a truncated-shifted Lennard-Jones potential \cite{1971-weeks}, 
\begin{equation}
  \label{tsLJ}
V_{tsLJ} = \left\{ \begin{array}{ll} 
     U_{LJ}(r) -U_{LJ}(\rcut) ,& \mbox{for $ r < \rcut$ }  \\ 
                                        0 ,& \mbox{for $ r  \geq  \rcut$} 
\end{array} \right. ,  
\end{equation}  
where $U_{LJ}(r)=4 \epsilon [ (\sigma/r)^{12} - (\sigma/r)^6 ]$.
The $LJ$ energy parameter $\epsilon$, is given in units of the experimental well depth $\epsilon_e$, and any energy $U$ will be also referred to $\epsilon_e$, \ie $U=U_e/\epsilon_e$. In the same way we choose the Boltzmann constant to be $k=1$ in reduced units, and therefore the reduced temperature is $T=k_e T_e/\epsilon_e$. The modulus of the dipole moments can be also expressed in the reduced system  as: $\mmu = \mu^2_e/(4 \pi \mu_{o,e} \sigma^3_e\epsilon_e)$. It should be noted that the soft-core and the attractive part have been implemented as in eq. (\ref{Ushort}) and not through a simple $LJ$ potential because we want to ensure that the effective repulsion is roughly the same when different values of  $\epsilon <1$ are used. In this way a comparison between chains with different depths for the attractive well and  non-sticky chains can be performed more easily.

The colloidal particles are assumed to interact pair-wise as point dipoles according to the potential,
\begin{equation}
\label{dipdip} 
U_{dip}(\br_{ij}) =   \frac{\bmu_i \cdot \bmu_j }{|\br_{ij}|^3} - \frac{3[\bmu_i
\cdot \br_{ij}][\bmu_j \cdot \br_{ij} ]}{|\br_{ij}|^5} ,
\end{equation}
where ${\br}_{ij} = {\br}_i - {\br}_j$ is the displacement vector between particles
$i$ and $j$. The energy due to magnetic interactions is calculated by direct summation over all pairs of particles. In spite of being algorithmically ${\cal O}(N^ 2)$, for small numbers of colloids this is the fastest and most accurate way to compute it when open boundary conditions are used (see below). Reasonable values of  $\mu=|\bmu|$ depend in general on the composition and size of the colloidal particles. Aside from cobalt nanoparticles, colloidal particles found in common commercial ferrofluids usually do not exceed values of $\mu \sim 10$.

\begin{figure}
\begin{center}
\subfigure[]{\label{fig:f1a}\includegraphics*[width=4.2cm]{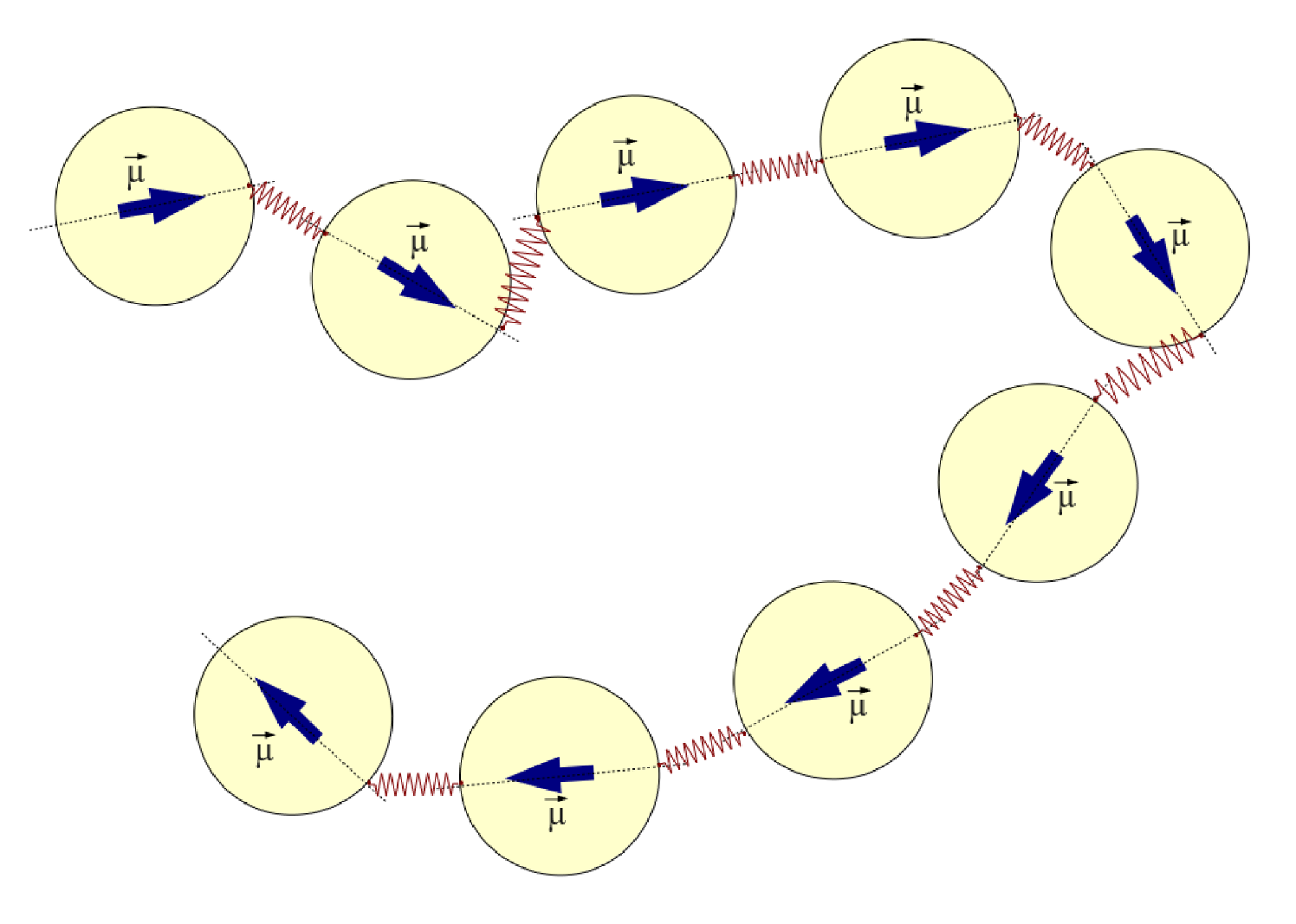}}
\subfigure[]{\label{fig:f1b}\includegraphics*[width=4.2cm]{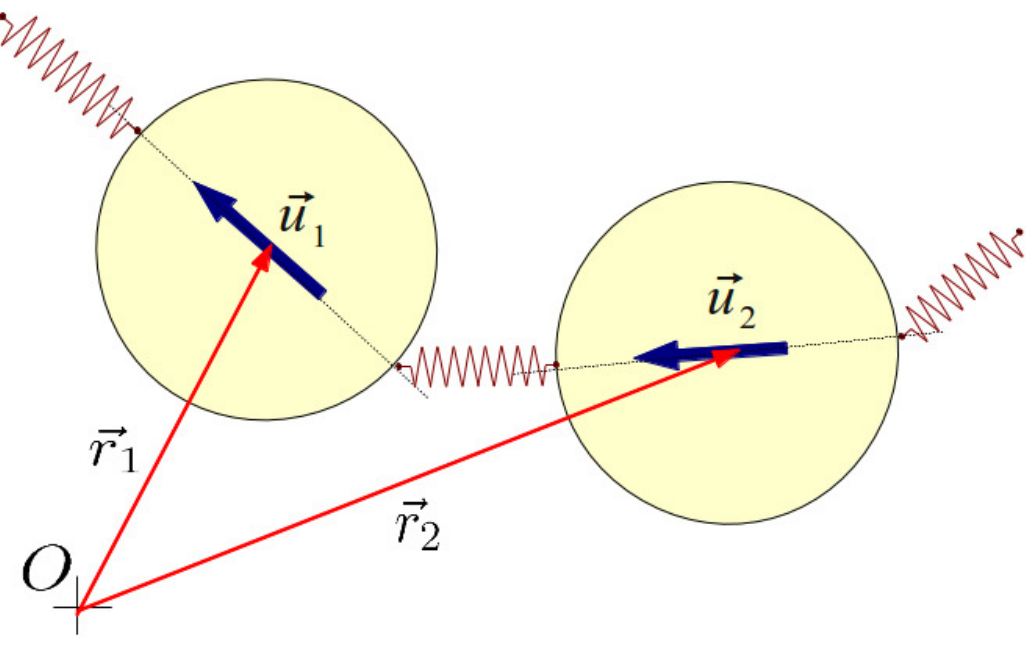}}
\caption{The magnetic filament is modelled as a chain of beads linked by springs anchored onto a point of their surfaces, see eq. (\ref{SOSpotencial}). The magnetic moment of each particle is used as reference vector to define the unitary vectors ${\bf \hat u} = {\bmu} / \mu$ which are used to determine the position of the anchoring points on the surface of the particles, $\sigma/2~\hat{\bet}$ and $-\sigma/2~\hat{\bet}$, when the centre of the particle is taken as the origin.}
\label{f1:plotmodel}
\end{center}
\end{figure}

In order to connect the colloids to form a chain, a linking model in which springs between consecutive particles are not anchored at the centres of the beads, but at fixed points on their surface, has been implemented. 
This was done in order to effectively penalise those conformations in which consecutive particles, holding a magnetic mono domain blocked in the crystal structure, are in an inverted orientation, resulting in an anormal stretching of the bond.
The model is shown schematically in Figure \ref{f1:plotmodel}. The proposed spring potential is written as
 \begin{equation}
     U_{s}(\br_i,\br_j,\hat{\bet}_i,\hat{\bet}_j) = \frac{1}{2} K_{S} \left(\br_i - \br_j - (\hat{\bet}_i+\hat{\bet}_j) \frac{\sigma}{2} \right)^2,
     \label{SOSpotencial}
  \end{equation}
where $\br_i$ and $\br_j$ are the position vectors of the centres of the beads. $\hat{\bet}_i$ and $\hat{\bet}_j$ are unitary vectors placed along the direction  that joins the two anchoring surface points of each bead (see Figure \ref{f1:plotmodel}). Thus, the anchoring points are collinear and  located at $\bl_i^{\prime}=\hat{\bet}_i\sigma/2$ and $\bl_i=-\hat{\bet}_i\sigma/2$ with respect to the centre of the bead. We assume all links in the chain are formed according to the following scheme: the point on the surface of the $i-1$ particle with position  $\br_{i-1}+\bl_{i-1}^{\prime}$ is linked to the point on the surface of the particle $i$ with position  $\br_i+\bl_{i}$. In order to penalise those conformations with consecutive dipoles in anti-parallel configuration, we associate each vector director  $\hat{\bet}_i$ with  the dipole moment of the particle, $\bmu_i$, i.e. $\hat{\bet}_i=\bmu_i/|\bmu_i|$. The constant of the potential is set to $K_{s}=30$ which is enough to ensure the average bond length to lie within a reasonable range $r_{bond} \in [0.98,~1.1] \sigma$. The use of larger values for $K_{s}$ is possible but 
implies a further reduction of the integration time step.

The numerical simulations are performed using Langevin dynamics, in which colloidal particles are moved according to the translational and rotational Langevin equations of motion that for a given particle $i$ are\cite{1987-allen}:
\begin{eqnarray}
M_i \frac{d{\mathbf v}_i}{dt} = {\bF}_i - \Gamma_T {\bv}_i + {\bxi}^T_i \\
{\bI}_i \cdot \frac{d{\bomega}_i}{dt} = {\btau}_i - \Gamma_R {\bomega}_i + {\bxi}^R_i
\end{eqnarray}
where $\bF_i$, and $\btau_i$ are respectively the total force and torque acting on the particle $i$. $M_i$ and $\bI_i$ are its mass and inertia tensor, and $\Gamma_T$ and $\Gamma_R$ are the translational and rotational friction constants. $\bxi_i^T$ and $\bxi_i^R$ are Gaussian random forces and torques, each of zero mean and satisfying the usual fluctuation-dissipation relations. In the simulations, $t=t_e\sqrt{\epsilon_e/(m_e\sigma_e^2)}$, where $m_e$ is the real mass of the colloids;  $F=F_e \sigma_e/\epsilon_e$, and $\tau = \tau_e /\epsilon_e$. For equilibrium simulations, the values of the mass, the inertia tensor, as well as friction constants $\Gamma_T$, and $\Gamma_R$ are irrelevant because the same equilibrium state is  reached independently of their value. Only the dynamics to attain such equilibrium state may show differences. For simplicity, the particle mass is chosen to be $m = 1$ and we take the inertia tensor to be the identity matrix in order to ensure isotropic rotations $\bI = {\mathbf 1}$. We have chosen $\Gamma_T=1$ and $\Gamma_R=3/4$ because we observed that these values produced a conveniently fast relaxation to equilibrium\cite{2008-cerda-jp,2008-kantorovich}. The reduced time step is set to $\delta t = 5 \cdot 10^{-4}$ in order to ensure a correct integration of the equations of motion.

The simulation starts by placing the filament in an open three-dimensional non-bounded space with the position of the first bead located randomly. The remaining monomers are positioned using a self-avoiding random walk scheme with an overlap radius of $0.9\sigma$. The chain is pre-equilibrated at $T=1$ for $2 \cdot 10^5$ integrations with the magnetic interaction turned off while the time step is slowly increased from $10^{-3} \delta t$ till $0.05 \delta t$. Subsequently, magnetic interactions are turned on, and a second pre-equilibration stage consisting of $5 \cdot 10^5$ integrations is performed while gradually raising the time step from $0.1 \cdot 10^{-3} \delta t$ till $\delta t$. Right after, if the final temperature is $T<1$ we perform an annealing process using the final time step $\delta t$: the temperature is reduced from $T=1$ down to its final value by performing a set of five annealing stages of $5\cdot 10^5$ steps each one. Once the final temperature has been reached, the chain is equilibrated for a period of $3\cdot 10^6~e^{1/T} ~\delta t$ in order to ensure that the chain is in the thermodynamic equilibrium. After the equilibration period, the system is sampled at intervals of $2500~e^{1/T}~\delta t$ for another period of $15\cdot 10^6~e^{1/T}~\delta t$ to make sure there are no correlations between measurements. To get further assurance that the results do not depend on the initial conditions and to improve statistics, an average over $15$ independent runs for each set of sampled parameters $(T,\epsilon,\mu)$ is performed. The simulations have been performed using the package \es~\cite{2006-limbach}.

\section{Results and discussion}
\label{sec3}
In the present model there are two main competing interactions: on the one hand there is the $LJ$ attractive interactions among the beads that  tend to collapse the chain when the temperature is lowered\cite{2008-binder,2003-baysal} and, on
the other hand, there is the magnetic interaction which is known from ferrofluid studies\cite{2008-cerda-jp,2008-kantorovich} to favour the formation of  rod-like chains and rings in which dipoles tend to align in a nose-tail conformation. Nose-tail conformations are those in which two dipole $i$ and $j$ satisfy $\bmu_i \cdot \bmu_j = \mu^2$, and $\bmu_i \cdot \br_{ij} = \bmu_j \cdot \br_{ij} = \pm \mu~r_{ij}$. This noise-tail alignment allows to minimise the magnetic energy in eq.(\ref{dipdip}).  We define a dimensionless parameter
\begin{equation}
\eta \equiv  \frac{\epsilon\sigma^3}{\mu^2}
\end{equation}
which measures the relative strength of the attractive LJ interaction with respect to the strength of the magnetic interaction for particles at close contact, $r_{ij}=\sigma$, in a nose-tail conformation. 
 
Although we just focus on the behaviour of a single chain,  the number of parameters involved to explore the full phase diagram of a Stockmayer polymer is large. Therefore, in order to get a first sketch on the phase diagram, we have chosen to focus  in the case of a polymer with a fixed number of beads $N=100$. In addition, all beads are assumed to have the same magnetic moment  $\mu^2=5$ and  share the same $LJ$ energy parameter $\epsilon$. In this first approach no external magnetic field is present. The phase diagram will be studied as a function of two parameters: $\eta$, that is controlled modifying the value of $\epsilon$ while keeping $\mu$ fixed, and the temperature $T$. The range of values explored for the parameter $\eta$ is $\eta \in [0,0.2]$, where $\eta=0$ corresponds to the case of non-sticky chains explored in previous studies \cite{2011-sanchez,2013-sanchez}. The range of temperatures sampled is $T\in [0.27,5]$, where the upper boundary $T=5$ has been chosen because the most interesting features were found in all cases to occur well below such value. It is worth to remark that the value of the parameters chosen for the simulations correspond to values close to those one can expect in experiments, thus, for instance,  the relative strength of the magnetic forces involved compared to the thermal fluctuations  coincide with that corresponding to ferrofluid particles made of magnetite with $\sigma_e \sim 20-25nm$. 

Values of $\eta > 0.2$ and $T<0.27$ correspond to regions in the phase space characterised by compact structures. An effective sampling of these regions with usual Langevin methods is very costly in computer terms and requires special techniques like \eg umbrella sampling\cite{1987-allen}, or Wang-Landau\cite{2006-parsons,2009-seaton}, or other advanced existing methods\cite{2007-vogel,2009-schnabel-cpl}.

In the next sections we will characterise the typical conformational states of a Stockmayer polymer in the weak  ($\eta < 0.10$) and strong  ($\eta \ge 0.10$) attractive regime. 

\subsection{Filaments in the weak attraction regime ($\eta < 0.10$)}
\label{results1aseccio}

The radius of gyration $R_g$ and the end-to-end distance $R_{ee}$ are two important observables that can be very useful in order to follow the structural changes of a magnetic filament. The end-to-end distance
is defined as
$R_{ee}=\langle \left( \vec{r}_1-\vec{r}_N  \right)^2 \rangle^{1/2}$
where $\langle ... \rangle$  denotes an average over all the 
sampled conformations of the chain. On the other hand, we can define the 
{\em gyration tensor} through their elements,
\begin{equation}
\label{gyrationtensordef}
R_{\alpha,\beta} =  \frac{1}{2N^2} \left\langle \sum_{i,j=1}^N (r_{i,\alpha} - r_{j,\alpha})
(r_{i,\beta} - r_{j,\beta}) \right\rangle
\end{equation}
where $\alpha$ and $\beta$ denotes the Cartesian components $x$, $y$, and $z$.
 The tensor can be represented  as
a diagonalisable $3 \times 3$ matrix with three eigenvalues or principal moments henceforth labelled as $\lambda^2_1>\lambda^2_2>\lambda^2_3$. The radius of gyration is  $R_g = \sqrt{\lambda^2_1 + \lambda^2_2 + \lambda^2_3}$.

\begin{figure}
\begin{center}
\includegraphics*[width=\myfigurewidth]{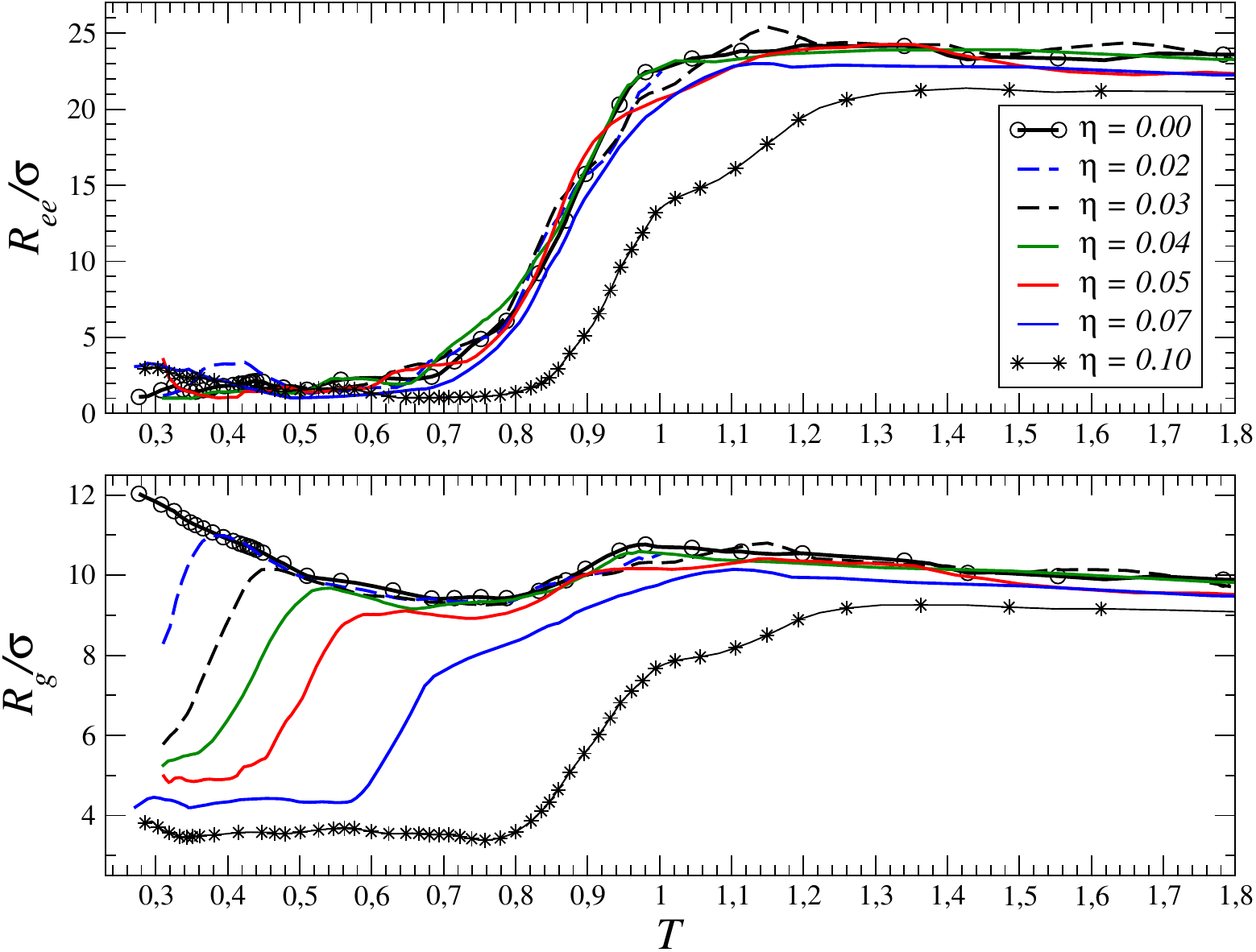}
\caption{The plot depicts the  the end-to-end distance $R_{ee}$ (top) and the radius of gyration $R_g$ (bottom) as a function of the temperature for $\eta<0.10$. The case $\eta=0.10$ has been also included for a better comparison with the results corresponding to $\eta>0.10$.}
\label{f2}
\end{center}
\end{figure}

Figure \ref{f2} shows the end-to-end distance (top) and the radius of gyration(bottom) as a function of the reduced temperature. Our results show that the ends of the chain tend to get closer to each other as the temperature is lowered. Remarkably, the behaviour of the end-to-end distance is very similar for all filaments with $\eta\in [0,0.07]$. In the range $\eta>0.07-0.10$  a noticeable two-fold decay step emerges whose origin is explained below. 

\begin{figure*}  
\begin{center}
\includegraphics*[width=\mypagewidth]{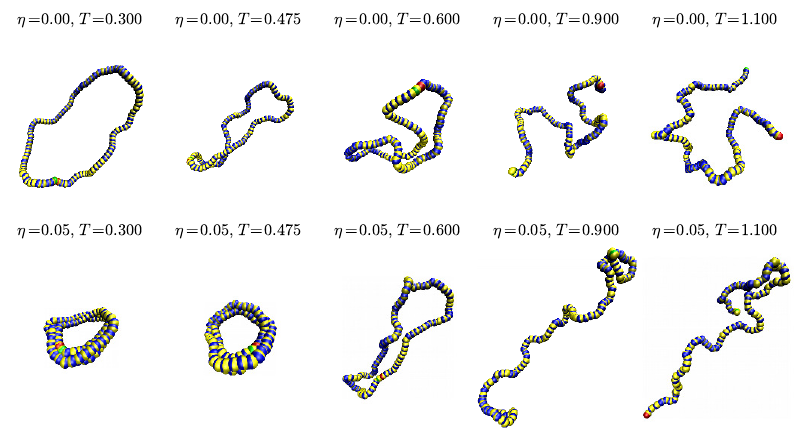}
\caption{Shown are several typical snapshots of the configurations that magnetic filaments of length $N=100$ adopt for different values of the temperature and relative strength of the attractive interactions $\eta$. Snapshots for $\eta=~0,~0.05$ are shown. The two ends of the filaments are show in green and red colours. The rest of the beads are painted in two colours (yellow and blue) to show the orientation of the magnetic moments of the particles.}
\label{f3}
\end{center}
\end{figure*}

The change in the radius of gyration also differentiates the behaviour  of the sticky filaments ($\eta>0$) from the non-sticky ones ($\eta=0$).  In the case of non-sticky particles the variation of $R_g$ is quite small. It ranges between $R_g \sim 9\sigma$ at $T=5$ (value not shown in Figure \ref{f2}) and $R_g \sim 12 \sigma$ at $T \sim 0.3$. This behaviour is clearly different from the one expected for non-magnetic chains: for such chains without attractive interactions, one would expect  $R_g$ to get a constant value corresponding to a self-avoiding walk. However, for a magnetic filament, as the temperature is lowered $R_g$  shows an initial expansion followed by a contraction in $T\in[0.7,1.0]$ and a second expansion at $T<0.7$. This particular behaviour is also observed in filaments with values of $\eta \rightarrow 0$ down to a temperature in which the attractive interactions dominate and induce a strong collapse of the chain.  This expansion-contraction behaviour can be understood as follows: as the temperature is lowered the magnetic interactions for an extended open chain dominate, favouring the stretching of the chain in a conformation in which all the dipoles tend to remain aligned. Nonetheless, below a certain temperature, the most favourable conformation is a closed structure because an extra aligned pair can be created by getting closer the ends of the chain, see Figure \ref{f3}. The entropic penalty of a closed chain is not excessive when the temperature is low enough. The closed structures at low temperatures are already known to occur in ferrofluids\cite{2008-prokopieva}, where particles assemble into ring-like structures. For low values of $\eta$ the transition from open to closed structures occurs at $T \sim 0.85$, which corresponds to a value of the dipolar coupling parameter of $\lambda \sim 6$. These results are in good agreement with those observed for non-sticky chains in the simpler variant of the bead-spring model\cite{2013-sanchez}.

\begin{figure}
\begin{center}
\includegraphics*[width=\myfigurewidth]{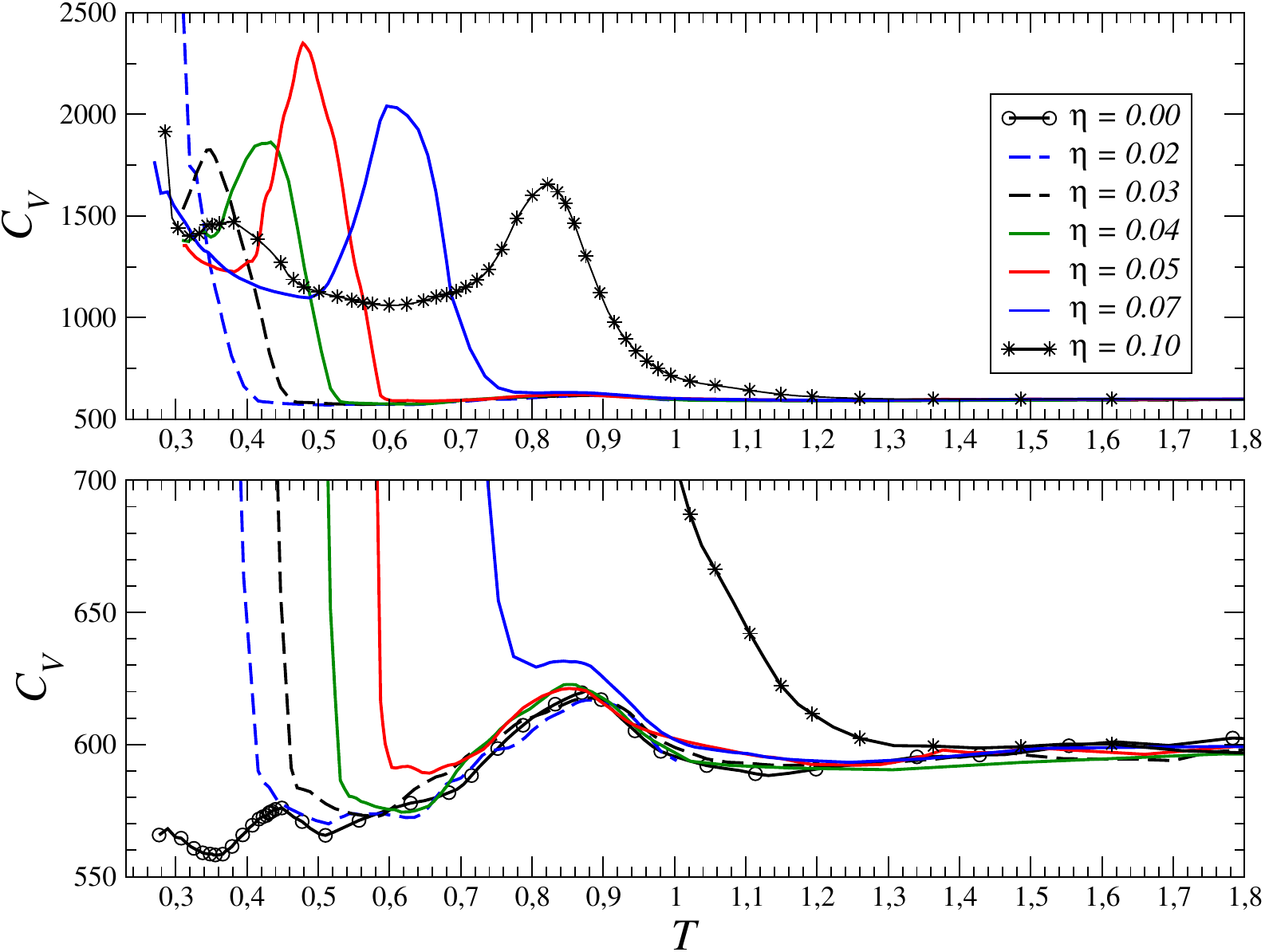}
\caption{The specific heat is represented as a function of temperature for different values of the relative strength of the $LJ$ attractive interactions $\eta$. The bottom plot is a zoom to highlight the small peaks that exist at $T \sim 0.8-0.9$ which point out the transition from extended open chains to simple closed structures.}
\label{f4}
\end{center}
\end{figure}

Further insight into the behaviour of the magnetic filament in the regime $\eta<0.10$ can be obtained by examining the specific heat $C_V$. Figure \ref{f4} shows the specific heat as a function of temperature. The bottom plot in Figure \ref{f4} shows some small peaks appearing around $T \sim 0.8-0.9$ which can be identified with the transition from open to closed structures. Two main features can be observed for such peaks: the first one is that for values of $\eta \in [0,0.07]$ they almost coincide in position and width. This means that this transition is almost independent of the strength of the attractive interaction, and is due to the absence of close contacts between particles which are not first nearest neighbours. 
The second feature is that the fluctuations in energy are small  compared to the peaks  observed in the top plot of Figure \ref{f4}. This fact is coherent with the idea that those peaks represent a transition from extended open to simple closed structures since the difference in the total energy should not be much larger than the energy originated in the creation of a new pair of aligned dipoles, plus the energy due to the close contact of two particles. These peaks in the specific heat are expected to grow with $N$ as in the case of non-magnetic homopolymers \cite{2005-rampf}, and non-sticky chains \cite{2011-sanchez}. 

\begin{figure}
\begin{center}
\subfigure[]{\label{fig:f5a}\includegraphics*[width=0.35\columnwidth]{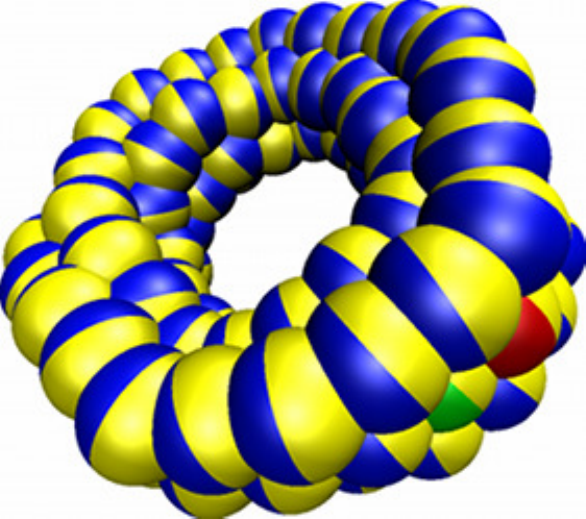}}\hspace{0.1\columnwidth}
\subfigure[]{\label{fig:f5b}\includegraphics*[width=0.35\columnwidth]{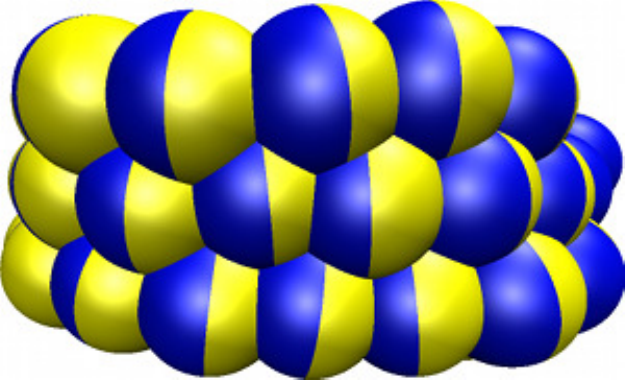}}
\caption{Displayed are two different perspectives of a typical conformation in the helicoidal state ($\eta=0.10$, and $T=0.475$). The colour code is the same as in  Figure \ref{f3}.}
\label{f5}
\end{center}
\end{figure}

The large peaks observed in the specific heat in the top of Figure \ref{f4} represent a different type of transition, namely, the conversion of simple closed structures into compact helicoidal-like ones as shown in Figure \ref{f3} (see snapshots at $\eta=0.05$ and $T<0.6$). As it is shown in Figure \ref{f5}, for the case $\eta=0.1$, the main part of the filament adopts a structure that resembles a tight helix, while the ends of the chain arrange in such a way that the two ends stay in close contact. These helicoidal states are related  to the toroidal conformations observed in non-magnetic semi-flexible chains where local chain stiffness helps to stabilise those structures\cite{2008-binder,2008-higuchi}. In our case, the magnetic interactions  tend to force a nose-tail orientation of the dipoles that, in addition, will induce a local chain stiffness\cite{2011-sanchez}. Nonetheless, there are some subtle differences between the toroidal conformations found in non-magnetic chains and the helicoidal structures observed here: in the non-magnetic case the toroidal walls are thick with a width of several particle diameters while in the magnetic case they tend to be much thinner. An open question is what would happen if longer chains of the order of $N \sim 10^3 - 10^4$ were studied. We argue that in the case of magnetic chains, a helicoidal state is preferred to a toroidal conformation for moderate values of  $\eta$. A helicoidal structure allows to minimise the energy associated to pairs of dipoles with their dipole moments lying parallel $\bmu_1=\bmu_2 \equiv \bmu$ but for which $\br_{12} \cdot \bmu =0$. Unlike the nose-tail conformations, these pairs have the highest possible magnetic energy and therefore are heavily penalised. Such energy penalty decreases as $\sim1/r^3$ with the distance between the two dipoles. In a toroidal conformation, a similar number of unfavourable dipole pairs may exist, but in difference to the helicoidal structure the distance between the two particles will be, in general, much shorter, leading to a higher energy penalty. The situation reverts for large values of $\eta$ in which the short ranged attractive interaction dominates over the magnetic one and a torus is preferred. Thus, helicoidal structures seem to be the result of a complex interplay between the attractive interactions that tend to collapse the chain into an isotropic globule, the magnetic forces which, on the one hand, tend to locally stretch the chain by leading to an effective local stiffness that favours toroidal conformations but, on the other hand, tend to avoid the formation of pairs of parallel dipoles with their relative vector position perpendicular to the direction of the dipoles and the chain entropy. 

Helicoidal states have also been found for non-magnetic chains when specific short-ranged square-well potentials were used \cite{2009-bannerman}. Another issue worth to mention is the possibility that those helicoidal states could be long-lived metastable transient states as those found by Sabeur et al.\cite{2008-sabeur} for simple homopolymers using truncated Lennard-Jones potentials. Sabeur et al. observed that the decay from the helical states is a stochastic rate-driven process, where the escape rate is $1/t_o \sim exp(\Delta E/kT)$ and $\Delta E$ is the height of the energy barrier. They  found a value of $t_o \approx 2000 \delta t$ for the particular case of a homopolymer chain of length $N=100$ at $T=0.04$. Since our equilibration and measurement times are of the order of $3 \cdot 10^6 ~e^{1/T}~\delta t$ and $15 \cdot 10^6 ~e^{1/T}~\delta t$ respectively, and taking into account that the lowest temperature we have sampled is of the order of  $T \sim 0.25$, we can reasonably conclude that our helicoidal structures are true equilibrium states. 

A quantitative way to characterise the formation of a helicoidal state is to evaluate an order parameter able to signal helix formation. Among the different order parameters\cite{2008-sabeur}, we have chosen the so-called $H_4$ parameter which characterises the global helical order defined as
\begin{equation}
\label{H4def}
H_4= \frac{1}{N-2} \left \langle \sum_{i=2}^{N-1} (\br_i-\br_{i-1}) \times (\br_{i+1}-\br_{i}) \right \rangle.
\end{equation}
$H_4=0$ is associated to isotropic conformations that resemble a rod, whereas $H_4=1$ holds for perfect helix. Other order parameters like $H_3$ \cite{1998-kemp} were found to lead to similar conclusions that those derived from $H_4$. 

\begin{figure}
\begin{center}
\includegraphics*[width=\myfigurewidth]{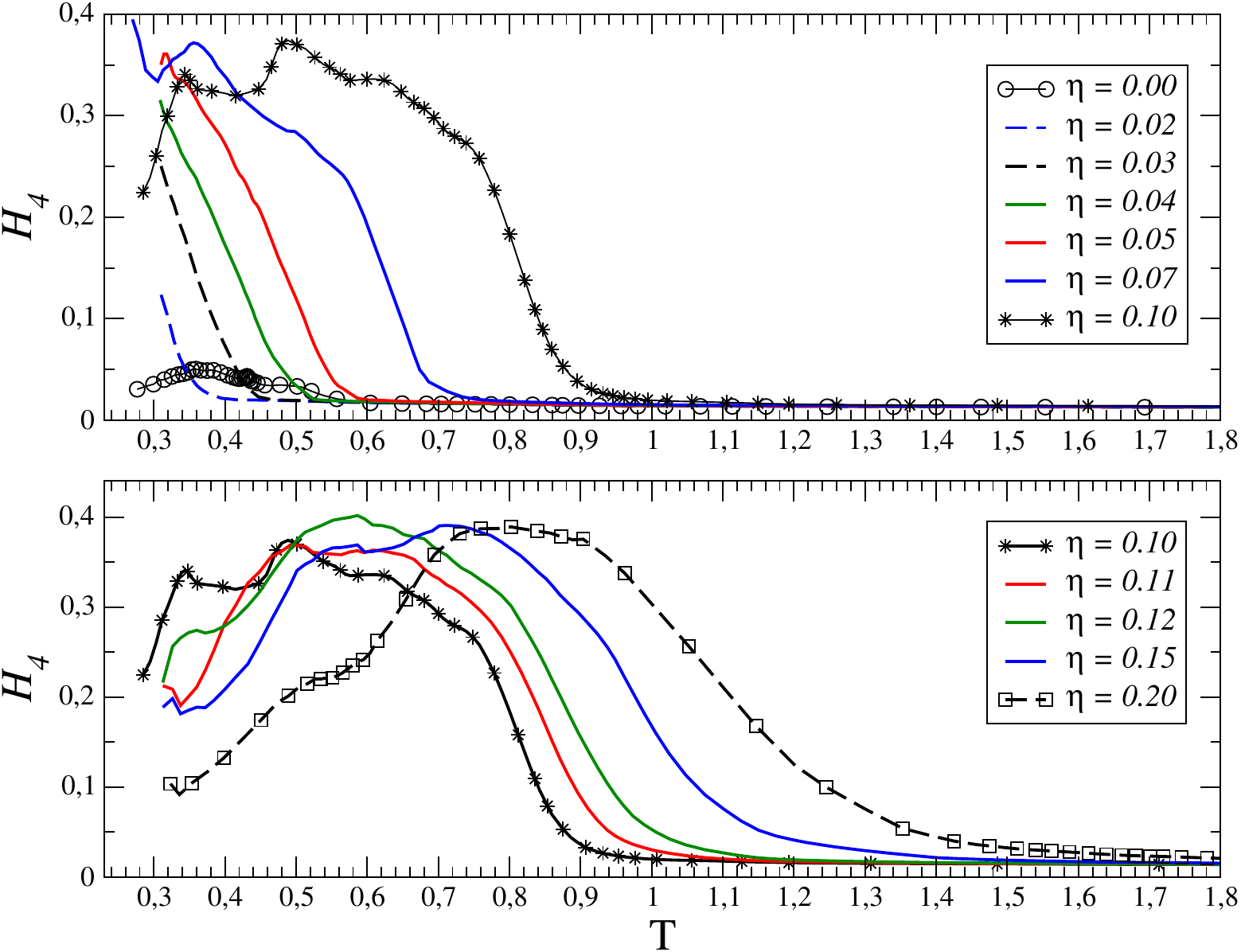}
\caption{The helicoidal order parameter $H_4$ as defined in eq.(\ref{H4def}) is plotted as a function of temperature for several values of $\eta \leq 0.10$ (top plot) and $\eta \geq 0.10$ (bottom plot).}
\label{f6}
\end{center}
\end{figure}

In Figure \ref{f6} (top plot) the value of the order parameter $H_4$ as a function of the temperature is depicted for $\eta \leq 0.10$. The values of $H_4$ remain equal to zero until temperatures close to the helicoidal transition point. The derivative, $dH_4/dT$, shows that the position of the inflection points occurring at the highest temperature coincides with the position of the peaks in the specific heat. This fact shows that both observables are linked to the appearance of helicoidal states. Figure \ref{f6} also displays that the achievement of the helicoidal states is very gradual: the further the temperature is lowered the higher is the value of $H_4$, and the structure looks more similar to a perfect helix. Only for values of $\eta \rightarrow 0.10$, the order parameter $H_4$ seems to reach a plateau within the range of temperatures sampled, and the largest values of $H_4 \sim 0.4$ are quite low compared with those of an ideal helix $H_4=1$. These relatively small values for the order parameter have a two-fold cause: the first one is that only the main part of the chain adopts a helix-like structure, whereas the ending parts of the chain arrange in such a way that the ends can be at close contact. If the whole structure was in a helicoidal state the two ends would be separated by a large distance, which gives an energy penalty. Such small differences in energy may be irrelevant at high temperatures but not at low temperatures. The second reason lies in the fact that these helicoidal states do not look like normal cylinders but rather they exhibit a symmetry similar to that of an elliptic cylinder, whereas the parameter $H_4$ is defined having in mind ideal helixes with a symmetry similar to that of a regular cylinder. Notice that the cross product  $(\br_i-\br_{i-1}) \times (\br_{i+1}-\br_{i})$ would be zero for a very elongated ellipse in which bonds between particles are locally aligned.

\begin{figure}
\begin{center}
\includegraphics*[width=\myfigurewidth]{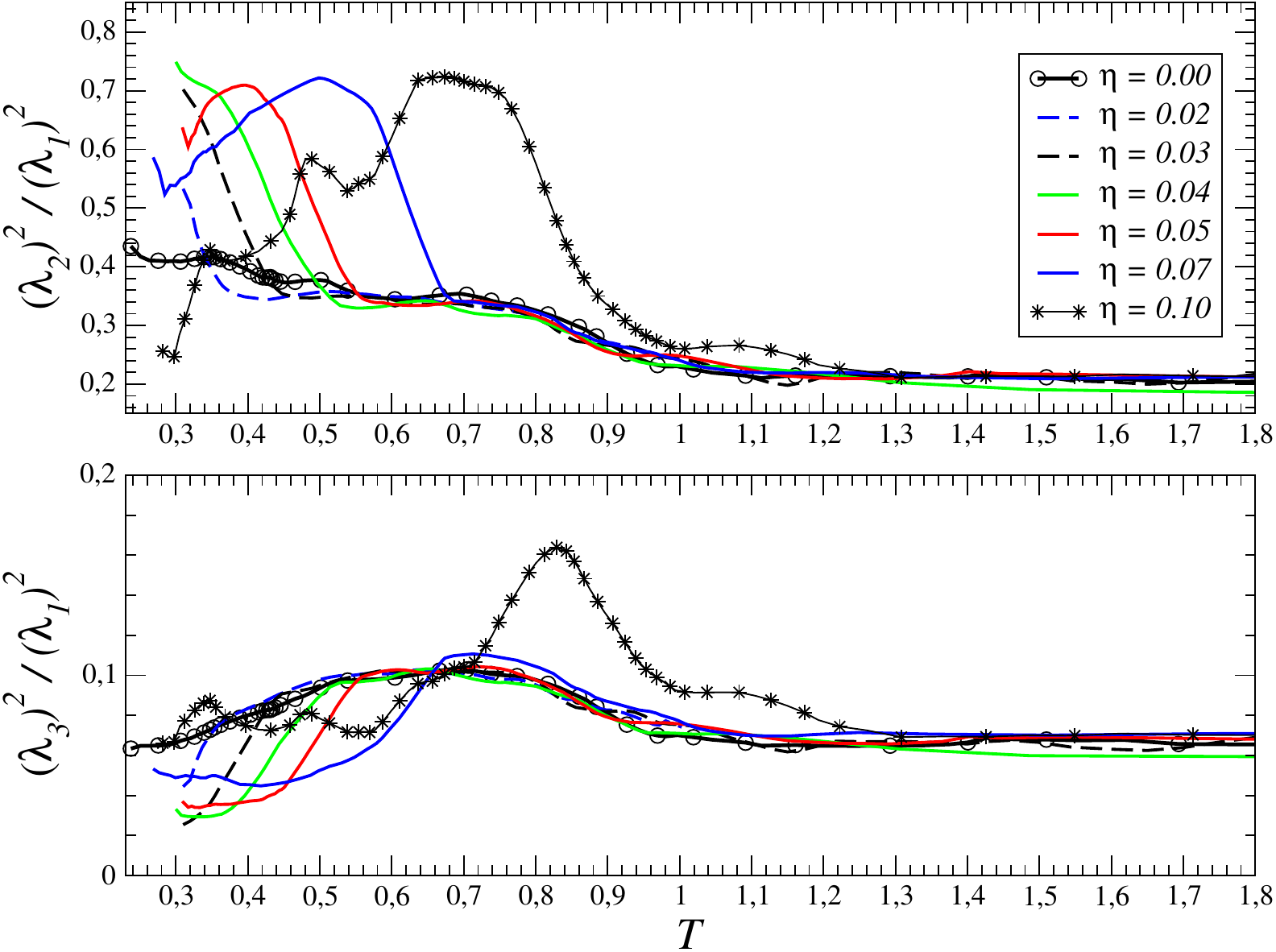}
\caption{The ratios of the second and third eigenvalues of the gyration tensor,  to the main eigenvalue are shown as a function of the temperature for several values of $\eta \leq 0.10$.}
\label{f7}
\end{center}
\end{figure}

The asymmetry of the filaments can be roughly inferred from a visual inspection of the different conformations shown in Figure \ref{f3}. However, a good set of observables to determine quantitatively the degree of asymmetry of the different conformations are the ratios of the second and third eigenvalues of the gyration tensor to the main eigenvalue, $\lambda^2_2/\lambda^2_1$ and $\lambda^2_3/\lambda^2_1$. Those ratios are shown in Figure \ref{f7} as a function of temperature. In the region of high temperatures $T \in [1.8,5]$  the ratios of the eigenvalues are approximately constant and equal to $\lambda^2_2/\lambda^2_1 \sim 0.2$  and $\lambda^2_3/\lambda^2_1 \sim 0.06$, respectively. These ratios mean that filaments are highly elongated along one direction and almost lie in a plane because the value of third eigenvalue is almost negligible compared to the first and second eigenvalues. These behaviour roughly corresponds to the tendency of the magnetic particles to align  in a row. The fact that the value of the ratios is almost the same for all $\eta \leq 0.10$ implies that in the range of high temperatures the attractive interaction plays a very minor role.  

At first sight, it can be surprising that all closed and collapsed structures do not show, in average, an isotropic distribution of particles, at least along their two main axis of rotation. From the isotropy of the space,  symmetric conformations are expected. However, the eigenvalues are obtained in the frame of reference of the particle ($frp$) and not in the frame of reference of the lab ($frl$). Due to the rotational averaging, the $frl$ is not suitable  to distinguish between a truely isotropic object from an anisotropic one. Since the average is taken over all possible rotations of the object, an anisotropic object  will always appear in average as isotropic in the $frl$. In fact, it is a well-stablished result that the averaged shape  (in the frp) of even a  fully flexible coil is not isotropic\cite{2007-alim,1971-solc,1934-kuhn}. The key to understand why a polymer chain prefers to adopt anisotropic conformations rather than isotropic ones is that the entropy is maximised for a single trajectory of the polymer when  the number of segments along each of the directions is inhomogeneous. Therefore, even in the limit of vanishing dipolar strenghts one should observe the shapes of the chains to be anisotropic.

\subsection{Filaments in the strong attraction regime  ($\eta \ge 0.10$)}
\label{results2aseccio}

\begin{figure}
\begin{center}
\includegraphics*[width=\myfigurewidth]{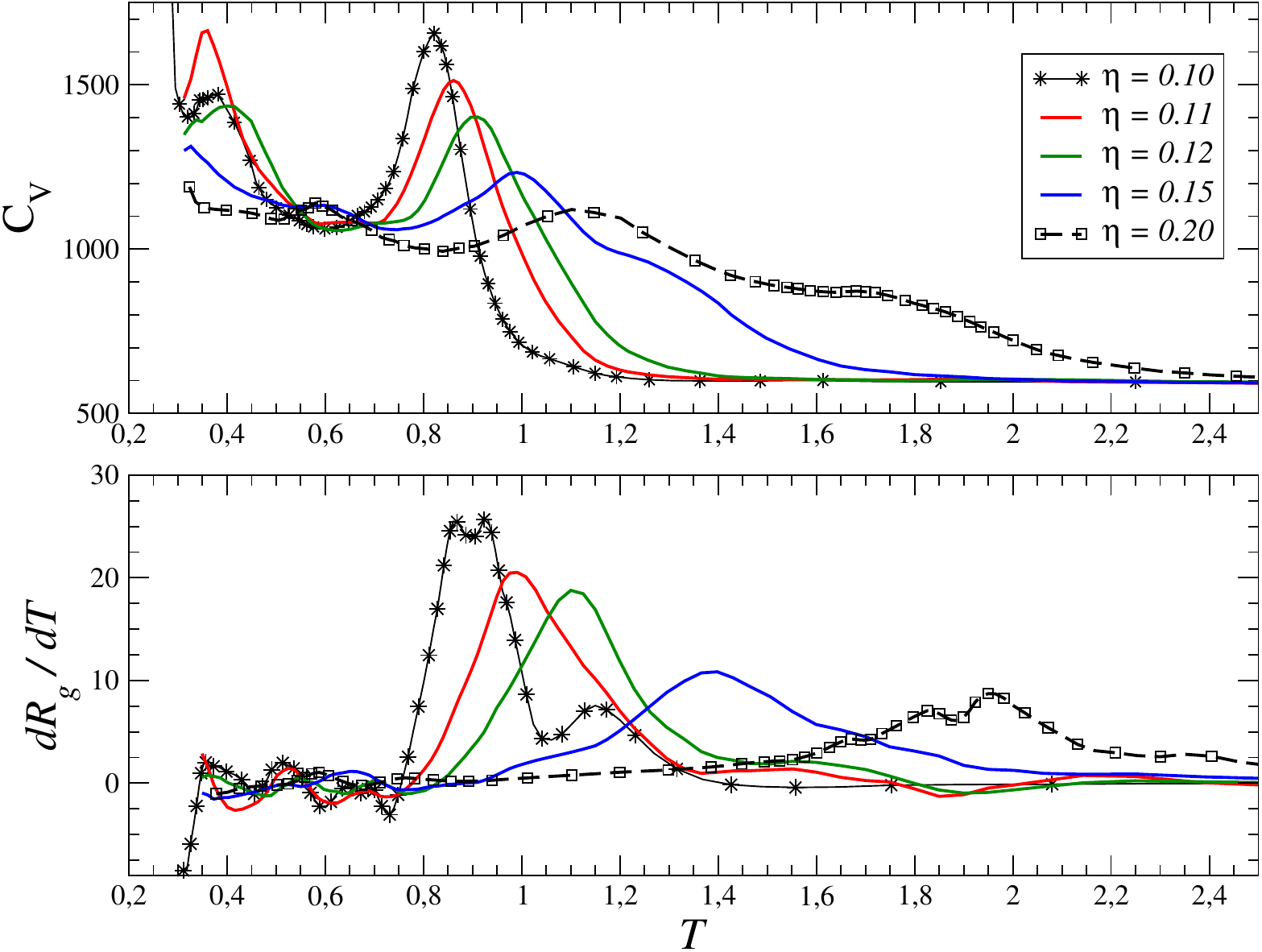}
\caption{The top plot shows the specific heat as a function of the temperature  for several values of the relative strength of the attractive interaction $\eta \ge 0.10$. The bottom plot displays the derivative of the gyration radius $dR_g/dT$ as a function of the temperature for the same range of $\eta$'s .}
\label{f8}
\end{center}
\end{figure}

\begin{figure}
\begin{center}
\includegraphics*[width=\myfigurewidth]{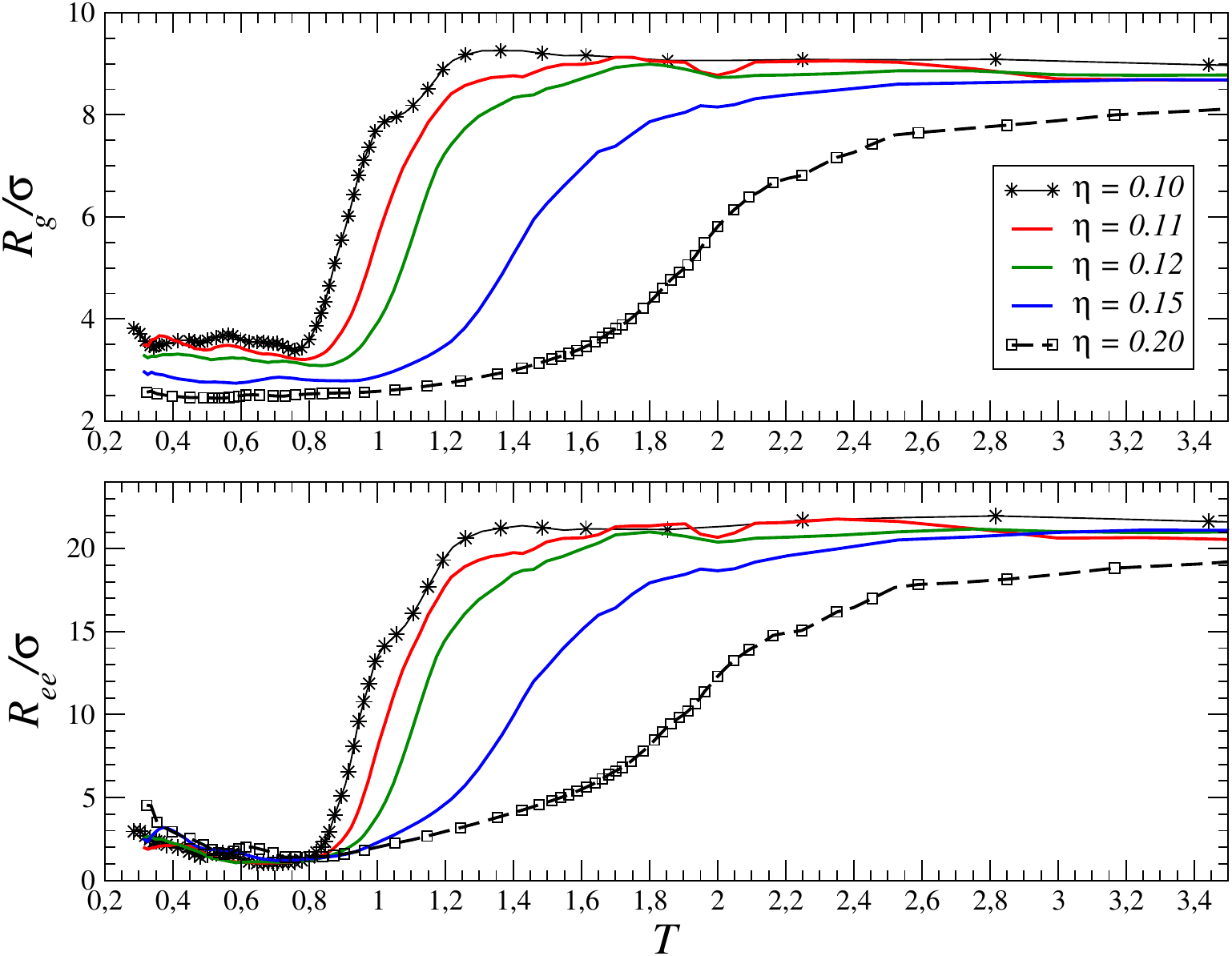}
\caption{The radius of gyration (top plot) and the end-to-end distance (bottom plot) are depicted as a function of temperature for several values of the relative strength of the attractive  to the magnetic interaction $\eta$. }
\label{f9}
\end{center}
\end{figure}

\begin{figure}
\begin{center}
\subfigure[]{\label{fig:f10a}\includegraphics*[width=0.27\columnwidth]{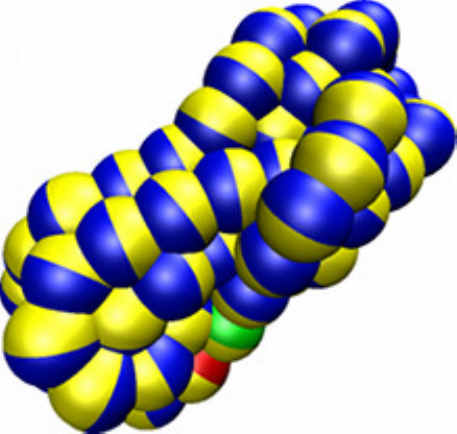}}\hspace{0.1\columnwidth}
\subfigure[]{\label{fig:f10b}\includegraphics*[width=0.49\columnwidth]{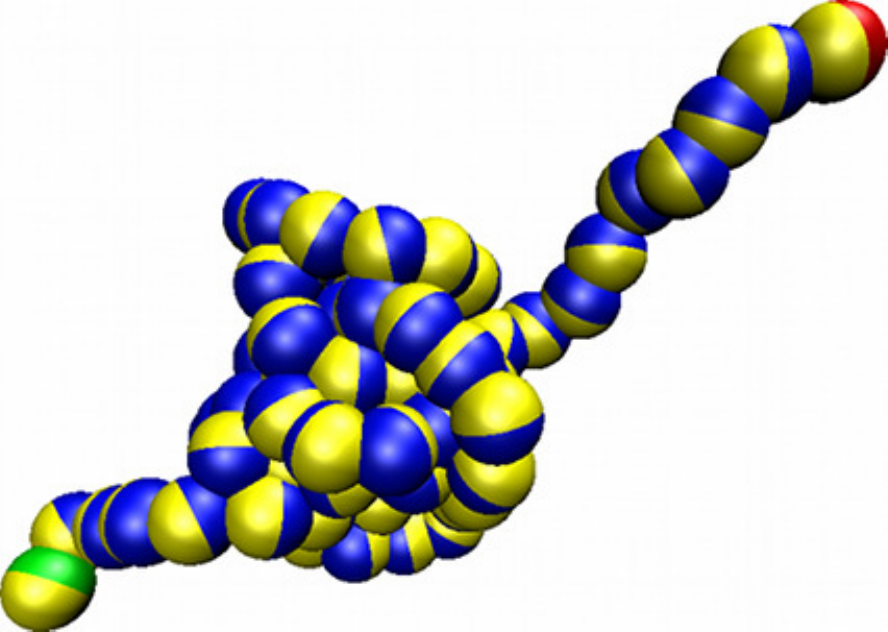}}
\caption{Displayed are two typical snapshots for high strengths of the LJ attractive interaction when compared to the strength of the magnetic interaction: (a) $\eta=0.15$, $T=0.40$. (b) $\eta=0.20$, $T=1.60$. The colour code is the same as in Figure \ref{f3}.}
\label{f10}
\end{center}
\end{figure}

We can observe in Figure \ref{f4} how the specific heat develops an additional peak for $\eta=0.1$ at $T \sim 0.4$. This peak is also observed for $\eta \ge 0.1$ (see top of Figure \ref{f8}) and, as $\eta \rightarrow 0.2$, it shifts towards higher temperatures $T \in [0.3,0.7]$ while decreasing in height. The radius of gyration and the end-to-end distance depicted in Figure \ref{f9} show that those peaks in the $C_V$ must correspond to a transition that takes place when the chain is already in a collapsed state. For that reason it is very difficult to discern from the structural parameters any relevant signal of the transition. This transition must involve an internal rearrangement of the particles without noticeable changes in the global size of the structure. A rough idea of such internal rearrangements is provided by the helical order parameter $H_4$ depicted in the bottom plot of Figure \ref{f6} for $\eta \ge 0.10$. A comparison between $C_V$ and $H_4$ (top of Figure \ref{f8}) reveals that the inflection point of the $H_4$ lines in the interval $T\in[0.3,0.7]$ roughly coincides with the position on the peak in the specific heat. We can infer that the new transition is associated with a loss of the helicoidal order and the onset of compact disordered states. A typical structure in this regime is represented in Figure \ref{f10}a.

In those collapsed states the  magnetic filament occupies a volume that is approximately $1.5$ times  the volume  of a compact hexagonal packaging (hcp) of $N=100$ spheres. This result shows that the magnetic filaments at $\eta=0.2$ and $T<0.6$ exhibit a substantial  degree of compaction. On the other hand, for $\eta>0.1$ we observe that the helicoidal structures developed are far more  isotropic than for  $\eta<0.1$. Thus, for  $\eta=0.2$ the highest ratios of the eigenvalues of the radius of gyration  are $\lambda^2_2/\lambda^2_1 \sim 0.9$ and $\lambda^2_3/\lambda^2_1 \sim 0.6$ that imply a significant increase in the level of isotropy when compared to those values for $\eta<0.1$ plotted in Figure \ref{f7}.

From the comparison among $C_V$, $R_g$ and $R_{ee}$ for  $\eta \ge 0.1$ we can extract more useful information about the magnetic chains. A remarkable feature in Figure \ref{f9}  is that  $R_g$ and $R_{ee}$  have a very similar dependence with the temperature.  In this case,  the attractive interaction is strong enough to force compaction to occur  at higher temperatures than those at which filaments would  suffer the closing transition. Once the compaction of the chain occurs, the distance between the chain ends must substantially decrease, and this leads to the closing transition right  after the compaction.

Another remarkable feature that can be extracted from Figure \ref{f9}  is the observation that for the curves corresponding to $\eta=0.15$ and $\eta=0.20$ it is possible to clearly associate the inflection points in $R_g(T)$ with the two emerging peaks in the specific heat (Figure \ref{f8} top) at $T \sim 1.3-1.4$ and $T \sim 1.8-1.9$, respectively. This behaviour suggests the existence of a regime of intermediate states between the expanded chains and the compact helicoidal configurations.  For values of $\eta \in [0.10,0.15)$ one can infer that the peak is also present but hidden in the long tail associated to the appearance of helicoidal states that take place in the interval $T \in [0.8,~1.1]$. For these smaller values of $\eta$ the fingerprint of the transition towards this intermediate regime is found in the double stage decay of both $R_{ee}$ and $R_g$ (Figure \ref{f9}). A natural question that arises is what kind of conformations do exist in such intermediate region. The plots of $R_g$ and $R_{ee}$ in Figure \ref{f9} show that for those intermediate states, the chain is still far from being fully collapsed and there is still, on average, a long distance between the two ends of the chain. Snapshots of such intermediate states, as the one shown in Figure \ref{f10}b, confirm the previous suggestion: between expanded open chains and the compact helicoidal states there exists a region of partially collapsed filaments in which a part of the filament is already in a compact state while the other parts are still in an expanded conformation, with the chain ends being separated by a relatively long distance. 

The characterisation of this new transition from open expanded chains to partially collapsed states by using the specific heat is hard to be accomplished for $\eta<0.20$. We found that the best observable to determine the transition point are the $dR_g/dT$ curves\cite{note-1} (see bottom of Figure \ref{f8}). The maxima of the peaks of such function for $\eta=0.15, ~0.20$, coincide with the apparent position of the emerging peaks for the $C_V$ at the highest temperature. Results for the transition temperatures obtained via $dR_g/dT$ are shown in Figure \ref{f11} as solid blue triangles-up for all values of $\eta$.

Those partially collapsed structures resemble vaguely  the core-shell structures found in non-magnetic semiflexible attractive chains\cite{2008-higuchi}. Nonetheless, in the present case we have a core plus some loose tails rather than a shell surrounding a core. This behaviour comes from the anisotropic nature of the magnetic interactions which favour relative straight segments far from the core rather than the wrapping of the core by the non-collapsed part of the chain. It is not yet clear what the behaviour of the filament will be in the limit $N \rightarrow \infty$ when long non-collapsed segments may exist.

\subsection{The phase diagram for an isolated flexible Stockmayer polymer}
\label{results3aseccio}

\begin{figure}
\begin{center}
\includegraphics*[width=\myfigurewidth]{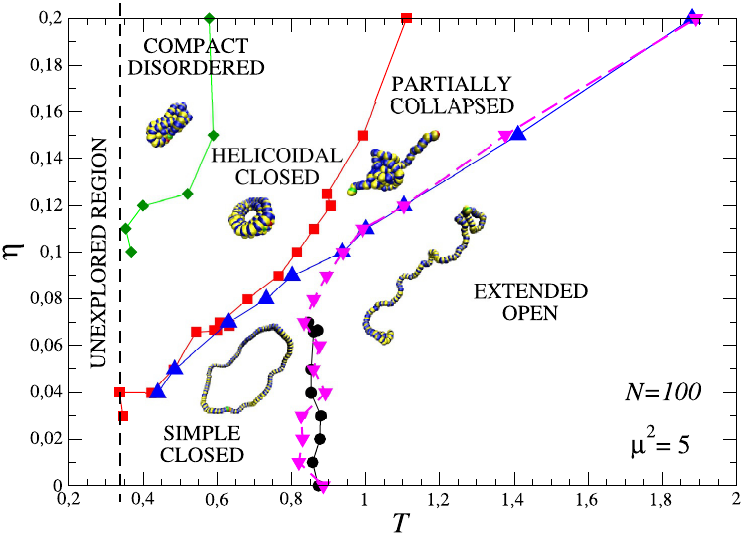}
\caption{A tentative phase diagram for magnetic filaments of length $N=100$ and $\mu^2=5$ is presented. See text in section \ref{results3aseccio} for a detailed explanation of how the different transition lines were obtained. In sake of clarity of the origin of each transition line, they have been painted using different colours and symbols.}
\label{f11}
\end{center}
\end{figure}

Gathering together the results presented in sections \ref{results1aseccio} and \ref{results2aseccio}, it is possible to build up a tentative sketch of the ($T$,$\eta$)-phase diagram as shown in Figure \ref{f11}.  The solid black circles correspond to the transition points for $\eta <0.10$ derived from the position of the maxima of the peaks in the $C_V$ and correspond to the transition from extended open chains to simple closed structures (see bottom  in Figure \ref{f4}). Solid red squares depict the transition points obtained from the maximum of the highest peaks of the $C_V$ in the range $T\in[0.7,1.2]$, see Figures \ref{f4} (top) and \ref{f8} (top). Those large peaks correspond to transitions towards a compact helicoidal state when the temperature is lowered. Solid green diamonds depict the transition points from compact helicoidal states to compact disordered states, which are obtained from the maxima of the peaks of the $C_V$ in the region of very low temperatures $T<0.7$ in Figure \ref{f8} (top). The solid blue triangles-up correspond to the maxima of the peaks in the  $dR_g/dT$, for $\eta \ge 0.1$, that also mark the inflection points of $R_g$, Figure \ref{f8}, (bottom). As we described in section \ref{results2aseccio} the position of such peaks should basically coincide with the transition points from open extended chains to partially collapsed states. Figure \ref{f11} also shows that $dR_g/dT$ gives a very good estimation of the transition points from simple closed chains to helicoidal states, in which transition temperatures derived from $dR_g/dT$ are very similar to those obtained from the position of the peaks in the $C_V$.  

In Figure \ref{f11} the solid black circles only refer to the transition for $\eta<0.7$. This is due to the fact that for higher values of $\eta$ the peak in the specific heat associated to such transition is hidden by the tail of the larger peak associated to the transition to helicoidal structures. In section \ref{results1aseccio} we mentioned the possibility of using the $dR_{ee}/dT$ in order to characterise the transition from extended open chains to simple closed states. The inflection points of the $R_{ee}(T)$ curves  are plotted in Figure \ref{f11} as solid magenta down triangles. Furthermore, for values of $\eta<0.1$ the inflection point in the end-to-end distance is clearly related to the transition point from open structures to simple closed structures. For values of $\eta \geq 0.1$ the inflection in the $R_{ee}$ takes place approximately at the same temperature as the inflection point of $R_g$ (solid blue triangles up) which, as discussed in section \ref{results1aseccio}, is a consequence of the fact that a partial compaction of the chain triggers the closing of the chain. 

A very remarkable fact observed in Figure \ref{f11} is the existence of two different conformational {\em'triple points'}. In the first 'triple point' extended open chains would coexist with partially collapsed states and simple closed states. In the second 'triple point', simple closed states will coexist with compact helicoidal states and partially collapsed states. The existence of two different, yet close triple points is a vivid example of how rich and complex the phase diagram is already for a single magnetic chain.  It is important to remark that in the different tests  performed,  no sign of hysteresis has been found for the transitions between collapsed phases, which further reinforces the idea that the structures found in this work correspond to equilibrium structures.

It is also interesting to discuss how the phase diagram would change if the strength of the dipolar interaction $\mu^2$ was modified, or if the range of the attractive forces was changed. To this end it is important to note that the bottom left corner of the phase diagram is dominated by the magnetic interactions, the upper left corner of the phase diagram is dominated by the LJ-like attractive interactions, and the bottom right corner is dominated by thermal motion. By increasing the dipolar strength $\mu^2$ while keeping the other factors unaltered we should expect the region where the magnetic interactions are dominant to expand. That means that the transition from extended open chains to simple closed structures should shift towards higher temperatures, and the transition from simple closed structures to compact helicoidal states should occur at higher values of $\eta$. The region where partially collapsed structures exist should also shift towards regions of higher values of $\eta$ and $T$. A reverse behaviour should be observed if we decrease the value of $\mu^2$ rather than increasing it. On the other hand, if we increase the range of the attractive interactions, the region where the attractive interactions dominate should expand, and therefore one should expect the transition from compact disordered structures to helicoidal states to occur at lower values of $\eta$. In turn, the transition from helicoidal to simple closed structures should also happen at lower values of $\eta$. It should be also possible to observe partially collapsed states at lower values of $\eta$.

Finally, it is also interesting to study the dependence of the phase diagram as function of the length  $N$ of the chain. In the limit $N \rightarrow 1$ it is clear that the helicoidal closed phase should not be present because the chain simply lacks sufficient monomers to form the helix. For the same reason, and based on the work of Jacobs-Bean\cite{1955-jacobs}, which is further supported by our experience with ferrofluids \cite{2008-cerda-jp,2008-kantorovich}, chains with $N<4$ are not expected to form loops. On the other hand
 Higuchi et al \cite{2008-higuchi} have  shown that for semiflexible non-magnetic chains, the transition from open to toroidal and partially collapsed structures occurs at higher temperatures as $N$ increases, and, due to the resemblances with our systems, we can expect a similar dependence with $N$ in our open-partially collapsed-helicoidal transitions. 
 
\begin{figure}
\begin{center}
\includegraphics*[width=\myfigurewidth]{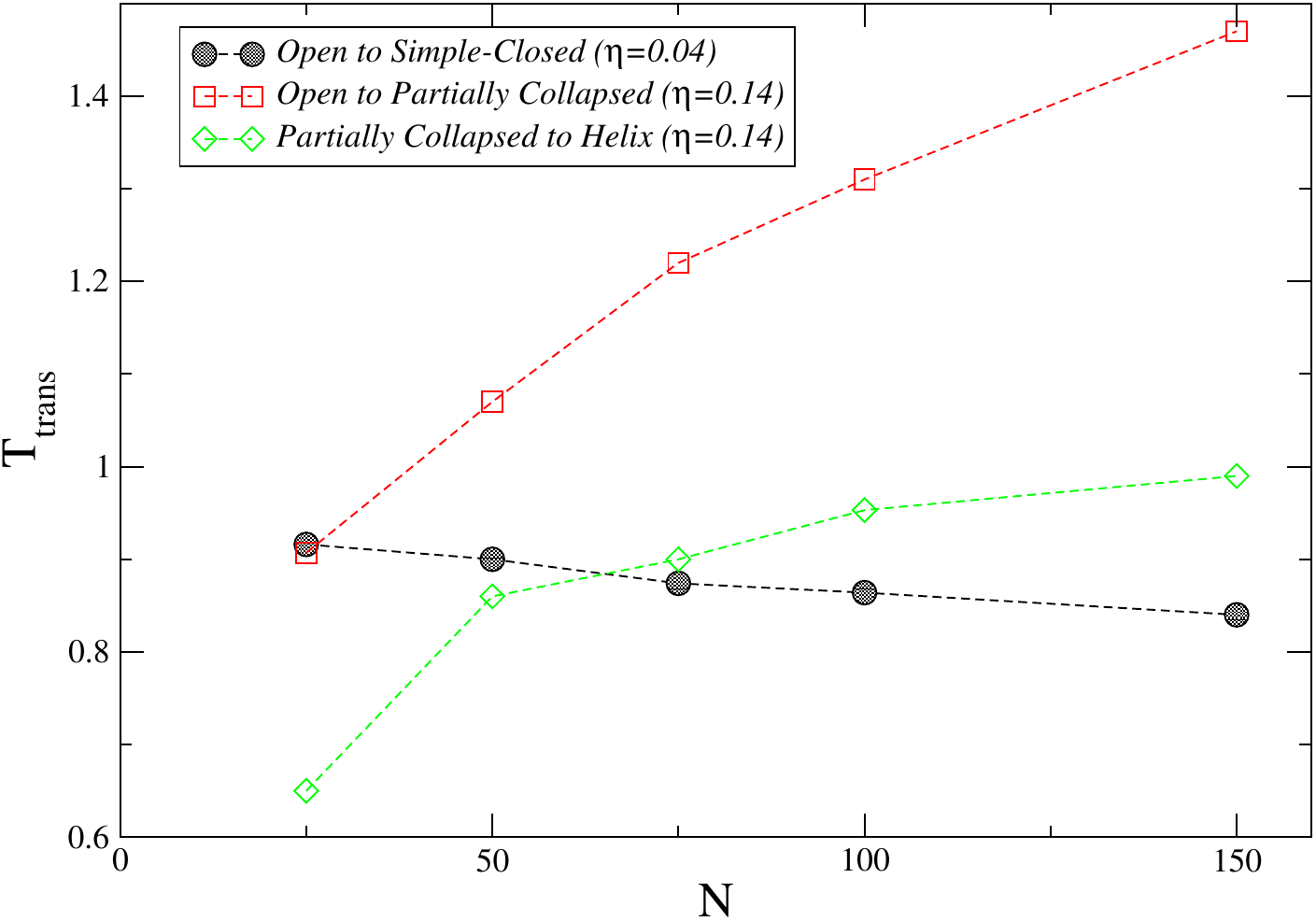}
\caption{The dependence of the characteristic transition temperature as a function of $N$ is shown for  the observed structural transitions from: open to simple closed (at $\eta=0.04$, black circles), open to partially-collapsed (at $\eta=0.14$, red squares), and partially-collapsed to helix transitions (at $\eta=0.14$, green diamonds).}
\label{f12}
\end{center}
\end{figure}

Figure \ref{f12} depicts the behaviour of the transition temperatures for the  open-closed, and the  open-partially collapsed-helicoidal transitions as a function of $N \in [25,150]$. Solid black circles show that for $\eta=0.04$ the open to simple-closed transition temperature decreases with $N$, i.e., longer chains in bulk  need further reductions in temperature to attain a closed shape. This behaviour has been also observed in the case of non-sticky  filaments ($\eta=0$) in bulk by Sanchez et al.\cite{2013-sanchez}. The behavior of the transition temperature as a function of chain length $N$ for the open to partially collapsed, and the partially collapsed to helicoidal transitions are represented in Figure \ref{f12} by the red squares and green diamonds, respectively. Our results  show an increase of the transition temperatures with $N$ that confirms our previous expectations based on the resemblances of the transitions we study in this work with those observed by Higuchi et al \cite{2008-higuchi}.

\section{Conclusions}
\label{sec4}

A tentative phase diagram  for a single Stockmayer polymer made of $N=100$ colloidal particles of identical size and magnetic moment $\mu$ is presented as a function of the temperature and the relative strength between the attractive LJ-like interactions versus the dipolar magnetic moment, $\eta$.  Our results, summarised in section \ref{results3aseccio}, evidence a rich phase diagram in which it has been possible to characterise up to five different conformational phases and two {\em 'triple points'}.

Although the present phase diagram is a simple sketch of a much more complex reality, several interesting open questions emerge from it. One of them is whether  it is possible to find a critical $\eta$ below which the closed-helicoidal transition vanishes.  The characterisation of the ground stated structures of the magnetic filaments and their comparison with the ground states observed in clusters of free Stockmayer-particles\cite{2005-miller} is also worthwhile to be studied. At intermediate temperatures, another challenging issue is to explore  whether compact helicoidal states will transit to compact globules directly or via intermediate partially collapsed states, where the existence of a third {\em 'triple point'} cannot be discarded. The changes in the phase diagram with colloidal size polydispersity, bond stiffness, and the presence of an external magnetic field, as well as a more elaborate study of the influences of the chain length $N$, are issues which need to be addressed in order to have a proper understanding of those systems. 

The knowledge of the different structures that a magnetic filament may adopt as a function of the interplay among the different interactions involved, and its conformational phase diagram, is crucial in order to assess the use of these filaments for new technological applications or as substitutes of current ferrofluids with enhanced properties. Magnetic filaments have an enormous potential for new applications, and the characterisation of their properties is still a pending issue. We expect the present work to constitute a first step towards the understanding of the magnetic filaments that  stimulates further developments on this subject of increasing scientific interest.

\section*{Acknowledgements}
Simulations were performed at the IFISC's Nuredduna high-throughput computing clusters, supported by the projects GRID-CSIC \cite{GRID-CSIC} and FISICOS (FIS2007-60327, funded by the Spanish MINCNN and the ERDF). We also thank the Junta de Andaluc\'ia  for its support via P11-FQM-7074 project (Spain).


\providecommand*{\mcitethebibliography}{\thebibliography}
\csname @ifundefined\endcsname{endmcitethebibliography}
{\let\endmcitethebibliography\endthebibliography}{}

\end{document}